\documentclass[aps,prl,twocolumn,superscriptaddress,showpacs]{revtex4-1}

\usepackage{amsmath}
\usepackage{amssymb}
\usepackage{graphicx}
\usepackage{epstopdf}
\newcommand{\sech}{\mathrm{sech}}

\newcommand{\tr}{\mathop{\mathrm{T}}}

\begin{document}

\title{Majorana Fermion Rides on a Magnetic Domain Wall}

\author{Se Kwon Kim}
\affiliation{
	Department of Physics and Astronomy,
	University of California,
	Los Angeles, California 90095, USA
}

\author{Sumanta Tewari}
\affiliation{
	Department of Physics and Astronomy,
	Clemson University,
	Clemson, South Carolina 29634, USA
}

\author{Yaroslav Tserkovnyak}
\affiliation{
	Department of Physics and Astronomy,
	University of California,
	Los Angeles, California 90095, USA
}

\begin{abstract}
We propose using a mobile magnetic domain wall as a host of zero-energy Majorana fermions in a spin-orbit coupled nanowire sandwiched by two ferromagnetic strips and deposited on an $s$-wave superconductor. The ability to control domain walls by thermal means allows to braid Majorana fermions nonintrusively, which obey non-Abelian statistics. The analytical solutions of Majorana fermions inside domain walls are obtained in the strong spin-orbit regime.
\end{abstract}


\maketitle

\emph{Introduction}.|Majorana fermions (MFs) have attracted much attention due to their non-Abelian exchange statistics, which makes them potentially useful for topological quantum computation \cite{*[][{, and references therein.}] NayakRMP2008}. In a film of a Rashba spin-orbit coupled (SOC) semiconductor subjected to a Zeeman field and proximity-induced $s$-wave superconductivity, a vortex in the superconducting order parameter supports one MF at its core \cite{*[{}] [{, and references therein.}] AliceaRPP2012,*[{}] [{, and references therein.}] StanescuJPCM2013}, as it does in a $p+i p$ superconductor \cite{IvanovPRL2001}. A one-dimensional wire with the same ingredients also supports MFs at the boundary between topological and nontopological regions. Such wires can form a mesh, where the MFs can be braided by slowly adjusting gate voltages, as proposed by \textcite{AliceaNP2011}.

Spintronics aims at the active control and manipulation of spin degrees of freedom in condensed-matter systems \cite{*[][{, and references therein.}] ZuticRMP2004}. Topologically stable magnetic textures, e.g., a domain wall (DW) in an easy-axis ferromagnetic wire or a vortex in an easy-plane ferromagnetic film, have been extensively studied out of fundamental interest as well as practical motivations exemplified by the racetrack memory \cite{ParkinScience2008}. Their dynamics can be driven by various means, e.g., an external magnetic field \cite{SchryerJAP1974}, an electric current (in metallic ferromagnets) \cite{SlonczewskiJMMM1996,*BergerPRB1996}, or a temperature gradient \cite{HinzkePRL2011,*YanPRL2011,*KovalevEPL2012, JiangPRL2013}. The topological phase transition of the SOC nanowire with proximity-induced superconductivity can be driven by adjusting the effective magnetic field \cite{OregPRL2010}. If we can  identify a localized magnetic texture that supports MFs under suitable conditions, we would, therefore, be able to manipulate them by controlling the magnetic host with standard spintronic techniques.

In this Letter, we propose to use a mobile DW in a proximate ferromagnet as a MF host. We show how to exchange MFs in the $Y$ junction by thermally-driven motion of the DWs, exhibiting non-Abelian braiding statistics. In conjunction with ongoing rapid developments in control and motion of magnetic DWs, it could be envisioned that networks of magnetic wires in proximity to a superconducting substrate can serve as robust and versatile means for storing and manipulating MFs. Specifically, we obtain analytic solutions for MFs inside a smooth DW in the strong spin-orbit regime. An intuitive idea of the existence of MFs inside the DW can be developed by casting the Bogoliubov-de Gennes (BdG) equation for MFs as two Jackiw-Rebbi Dirac equations with spatially-varying mass terms \cite{JackiwPRD1976}.

\begin{figure}
\includegraphics[width=0.9\columnwidth]{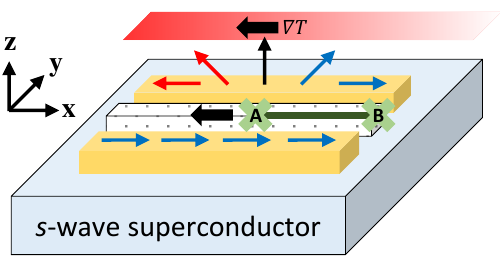}
\caption{(Color online) A schematic diagram of a device supporting two MFs in a SOC nanowire with proximity-induced exchange field and superconductivity. The wire is sandwiched between two easy-axis ferromagnets: one with a uniform magnetization and the other with a DW. In the right portion of the wire, where the magnetizations of the two ferromagnets are parallel, the net exchange field engenders the topological region, fomenting MFs at locations marked by A and B. The temperature-gradient induces magnon drag in the $x$ direction, which drives the DW and its accompanying MF at A to the hotter region.}
\label{fig:setup}
\end{figure}

\emph{MF at a magnetic DW.}|Consider a semiconducting nanowire with Rashba SOC, which is deposited on an $s$-wave superconductor and is proximity-coupled to adjacent ferromagnets \cite{StanescuJPCM2013,AliceaRPP2012}. See Fig.~\ref{fig:setup} for a schematic design of a device. The superconducting quasiparticle spectrum is obtained by solving the BdG equation $\mathcal{H}^\text{BdG} \Psi (x) = E \Psi (x)$, where
\begin{equation}
\mathcal{H}^\text{BdG} 	= \left( - \frac{\hbar^2}{2 m} \partial_x^2 - \mu + i \alpha \partial_x \sigma_2  \right) \tau_3
				+ \mathbf{M} \cdot \boldsymbol{\sigma} + \Delta \tau_1
\label{eq:H-BdG}
\end{equation}
acts on the spinor wavefunction $\Psi = \left( u_\uparrow, u_\downarrow, v_\downarrow, - v_\uparrow \right)^{\tr}$.
Here, $m$, $\mu$, and $\alpha$ are the electron effective mass, the chemical potential, and the strength of SOC. The proximity-induced exchange field $\mathbf{M}$ is perpendicular to the spin-orbit field $\propto\hat{\mathbf{y}}$. The proximity-induced superconducting order parameter $\Delta$ is gauge shifted to be real and positive. Pauli-matrix vectors $\boldsymbol{\sigma}$ and $\boldsymbol{\tau}$ act respectively on the spin and the electron-hole subspaces of the spinor $\Psi$. The corresponding quasiparticle creation operator is $\hat{\gamma}^\dagger = \int dx \sum_{\alpha = \uparrow, \downarrow}[ u_\alpha (x) \hat{\psi}^\dagger_\alpha (x) + v_\alpha (x) \hat{\psi}_\alpha (x)]$. For the uniform exchange field, the ``topological gap'' at zero momentum is given by
\begin{equation}
E_g = |\mathbf{M}| - \sqrt{\Delta^2 + \mu^2}.
\label{eq:E-g}
\end{equation}
When the gap $E_g$ is positive, $|\mathbf M| > \sqrt{\Delta^2 + \mu^2}$, the wire is in the topological phase, harboring a pair of MFs at its ends \cite{SatoPRL2009, *SauPRL2010,*AliceaPRB2010,*LutchynPRL2010,OregPRL2010}. Otherwise, for the negative gap, the wire is in the normal phase, without MFs.

A spatially-varying exchange field induces the topological phase transition along the wire where $|\mathbf M|$ crosses $\sqrt{\Delta^2 + \mu^2}$ \cite{OregPRL2010}. A DW in a ferromagnet adjacent to the wire is a natural object to bring about such a position-dependent field. We assume that the energy of the ferromagnet is given by
\begin{equation}
U[\mathbf m (x) ] = \int dx \left[ A |\partial_x \mathbf m|^2 - K_x m_x^2 + K_y m_y^2 \right] / 2,
\label{eq:U}
\end{equation}
where $\mathbf{m} (x)$ is the unit vector in the direction of local magnetization. Here, $A$, $K_x$, and $K_y$ are the positive coefficients characterizing the stiffness of the magnetization, the easy-axis anisotropy along the wire, and the hard-axis anisotropy in the spin-orbit field direction, respectively \cite{SchryerJAP1974}. Two ground states have uniform magnetization $\mathbf{m} \equiv \pm \hat{\mathbf{x}}$. A DW is a topological soliton solution minimizing the energy~(\ref{eq:U}) with the boundary condition $\mathbf{m} (x \rightarrow \pm \infty) = \pm \hat{\mathbf{x}}$: its magnetization is $\mathbf{m} (x) = \tanh(x / \lambda) \hat{\mathbf{x}} + \sech(x / \lambda) \hat{\mathbf{z}}$ where $\lambda = \sqrt{A / K_x}$ is the DW width \cite{SchryerJAP1974}. 

We sandwich the wire between two ferromagnets: one with a DW and the other with a uniform magnetization, as shown in Fig.~\ref{fig:setup}. The proximity-induced exchange field is described by
\begin{equation}
\mathbf{M} (x) = M_1 \left[ \tanh(x / \lambda) \hat{\mathbf{x}} + \sech(x / \lambda) \hat{\mathbf{z}} \right] + M_2 \hat{\mathbf{x}},
\label{eq:M}
\end{equation}
which introduces a new length scale $\lambda$ to the Hamiltonian. The presence of the DW causes spatial variance of the gap:
\begin{equation}
E_g (x) = \sqrt{M_1^2 + M_2^2 + 2 M_1 M_2 \tanh(x / \lambda)} - \sqrt{\Delta^2 + \mu^2}.
\end{equation}
For example, when $M_1 = M_2$ and $2 M_1 > \sqrt{\Delta^2 + \mu^2}$, there must be a pair of MFs in the wire: one at the right end, $x \rightarrow +\infty$, and the other one at the DW, $x_0 = \lambda \tanh^{-1} [ (\Delta^2 + \mu^2) / 2 M_1^2 - 1 ]$. The topological stability of the DW protects the hosted MF from disturbances of magnetization.

The technique to control a DW has been developed over decades in spintronics \cite{ZuticRMP2004}, which is directly translated into the ability to manipulate MFs. DWs are conventionally driven by an external magnetic field \cite{SchryerJAP1974} or a spin-polarized electrical current (in an itinerant ferromagnet) \cite{BergerPRB1996}. These methods, however, affect the electrons in the wire by altering the Hamiltonian. Instead, we propose to induce the motion of a DW by thermal drag \cite{HinzkePRL2011}, which is not intrusive to the electrons (see Fig.~\ref{fig:setup}). A temperature gradient generates a magnon current flowing to the colder region. Magnons adjust their spins toward the local direction of magnetization and react by exerting a torque on the DW. Addition of angular momentum to the DW causes its shift toward the hotter region. The resultant velocity of the DW, therefore, is proportional to temperature gradient. Multiple DWs on the single wire can be moved simultaneously, thereby providing the ability to control a series of MFs.

\begin{figure}
\includegraphics[width=0.3\columnwidth, page=1]{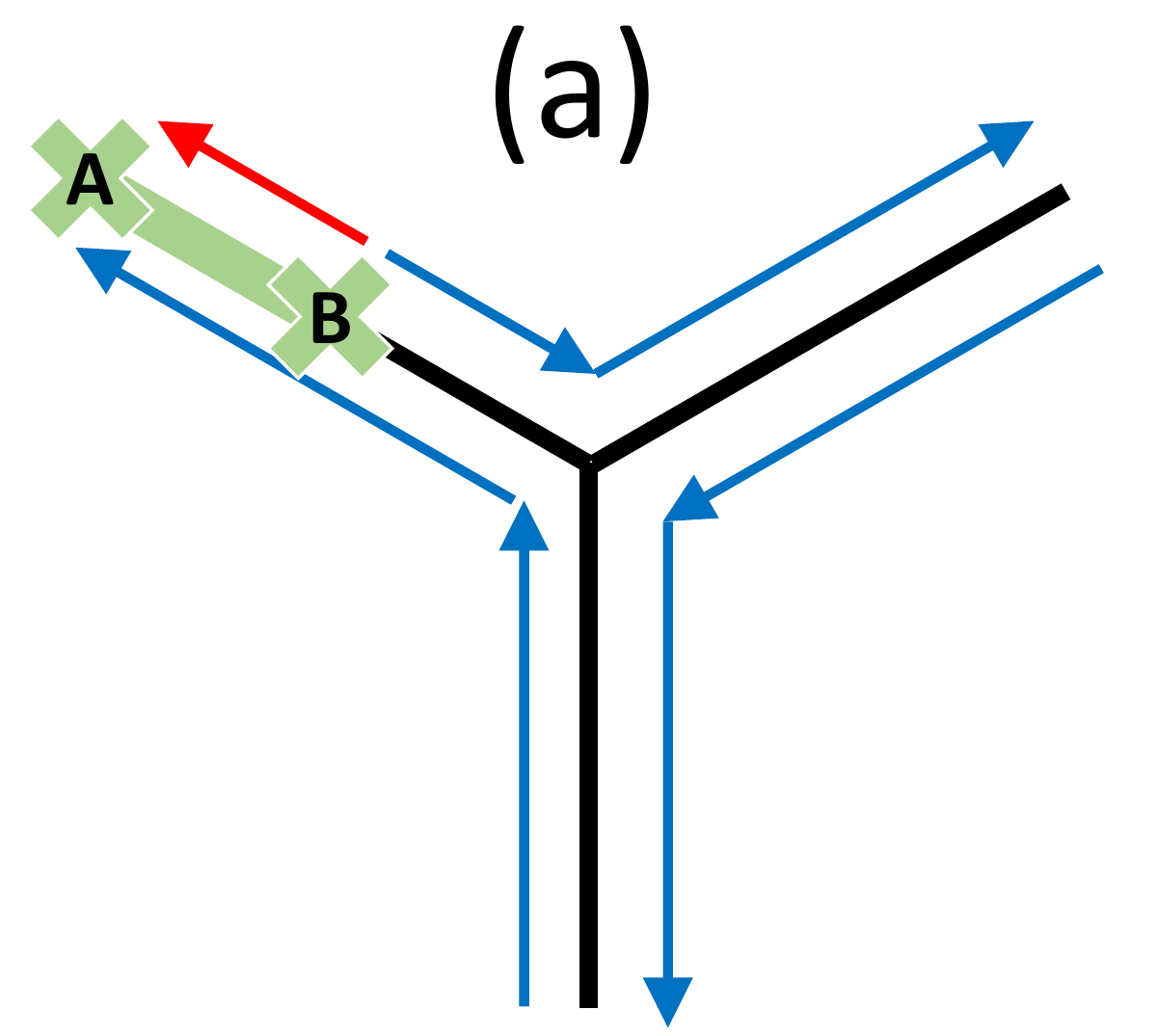}
\includegraphics[width=0.3\columnwidth, page=2]{fig2.pdf}
\includegraphics[width=0.3\columnwidth, page=3]{fig2.pdf}
\\
\vskip 0.2cm
\includegraphics[width=0.3\columnwidth, page=4]{fig2.pdf}
\includegraphics[width=0.3\columnwidth, page=5]{fig2.pdf}
\includegraphics[width=0.3\columnwidth, page=6]{fig2.pdf}
\\
\vskip 0.2cm
\includegraphics[width=0.3\columnwidth, page=7]{fig2.pdf}
\includegraphics[width=0.3\columnwidth, page=8]{fig2.pdf}
\includegraphics[width=0.3\columnwidth, page=9]{fig2.pdf}
\caption{(Color online) The exchange of two MFs bridged by a topological region performed by controlling the exchange field. (a)-(g) The positions of MFs can be exchanged by a series of DW displacements. (h), (i) Uniform rotation of the magnetization between MFs by $180^\circ$ about the spin-orbit field direction transforms the state (g) to the initial state (i). This is essentially monodomain flipping.}
\label{fig:exchange}
\end{figure}

\emph{Exchange of two MFs.}|The exchange of two MFs is possible in the Y junction of three nanowires, each of which is sandwiched between two ferromagnets. Figure~\ref{fig:exchange} shows the process to exchange MFs bridged by a topological region \footnote{One can also construct the process that exchanges two MFs bridged by a nontopological region, which is performed by domain wall displacements}. This proposal uses a well-controllable topological soliton (DW) as a host of MFs, and thus does not need intricate gate fabrication and control, unlike the exchange process proposed by \textcite{AliceaNP2011}. The non-Abelian exchange statistics of MFs is a universal property, meaning that it is invariant under the continuous deformation of the Hamiltonian as long as the gap $E_g (x)$~(\ref{eq:E-g}) remains finite, except at isolated locations of MFs \cite{AliceaNP2011, ClarkePRB2011, *HalperinPRB2012}. Our proposal uses the spatially varying exchange field $\mathbf{M} (\mathbf{r})$ and the chemical potential fixed at zero, $\mu(\mathbf{r})\equiv0$. These two functions, $\mathbf{M}(\mathbf{r})$ and $\mu(\mathbf{r})$, can be continuously deformed to a uniform exchange field in the $z$ direction and a spatially-varying chemical potential of the form studied by \textcite{AliceaNP2011}, without changing the gap. This is accomplished by first locally rotating the exchange field about the spin-orbit field direction to be oriented along the $z$ axis: $\mathbf{M}' (\mathbf{r}) = M(\mathbf{r}) \hat{\mathbf{z}}$, while keeping the gap $E_g (x)$ unchanged. The following transformation then connects our Hamiltonian ($s = 0$) to that of Ref.~\cite{AliceaNP2011} ($s = 1$):
\begin{equation}
\begin{split}
M(\mathbf{r}; s) &= M(\mathbf{r}) + \left[ - M(\mathbf{r}) + M_s \right] s, \\
\mu(\mathbf{r}; s) &= \sqrt{\left\{ \left[ - M(\mathbf{r}) + M_s \right] s + \Delta \right\}^2 - \Delta^2}.
\end{split}
\end{equation}
where $M_s=\max_\mathbf{r} M(\mathbf{r})$. The gap-preserving transformation between two exchange processes, one performed by controlling the exchange field and the other by gating the chemical potential, ensures that the process shown in Fig.~\ref{fig:exchange} exhibits the non-Abelian exchange statistics obtained in Ref.~\cite{AliceaNP2011}.

\begin{figure}
\includegraphics[width=0.4\columnwidth]{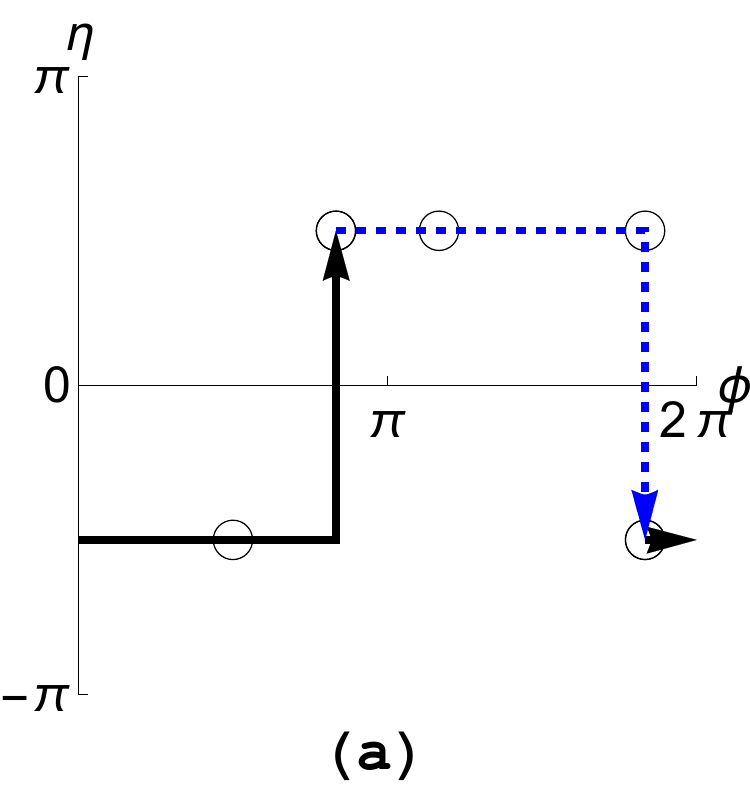}
\includegraphics[width=0.5\columnwidth]{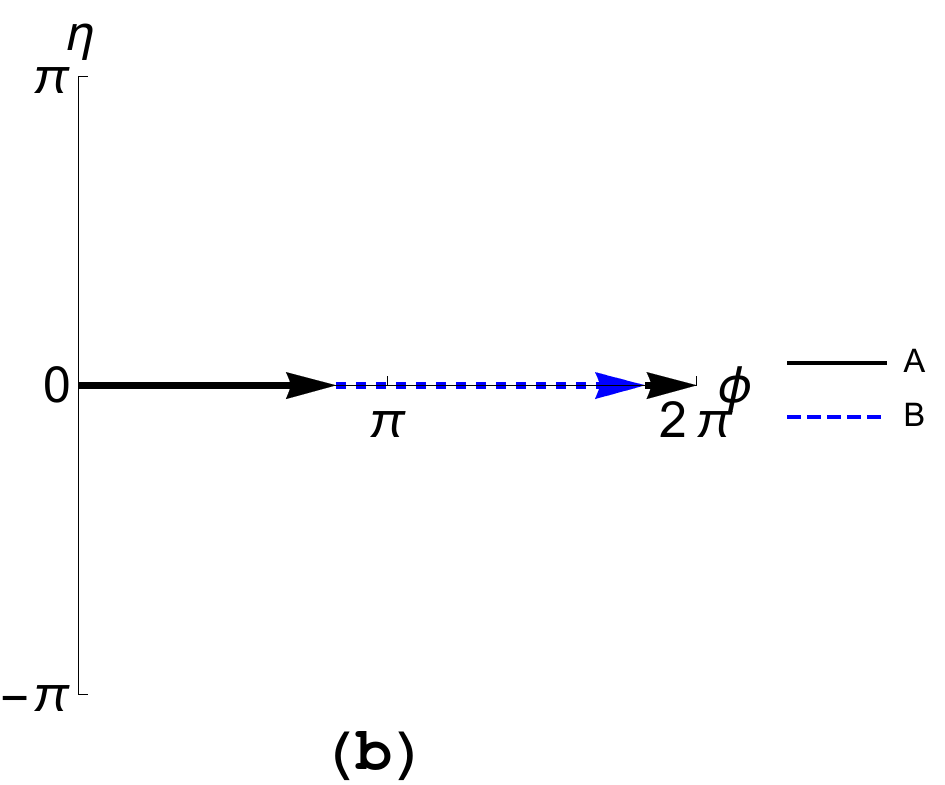}
\caption{(Color online) (a) The trajectories of the parameters $(\phi, \eta)$ of MF A and B in the proposed exchange process shown in Fig.~\ref{fig:exchange}. A circle shows the state of each step in the figure. The combined path of two trajectories forms a closed curve on the parameter space which is topologically a torus. (b) The trajectories of $(\phi, \eta)$ in the process proposed by \textcite{AliceaNP2011}. Two combined paths of (a) and (b) are topologically equivalent.}
\label{fig:parameter}
\end{figure}

Alternatively, we may understand the topological equivalence of two processes following the general approach developed by \textcite{HalperinPRB2012}. The state of each MF, denoted by A and B, can be described by two parameters. One is the wire orientation $\hat{\mathbf{w}}^a = ( \cos \phi^a, \sin \phi^a )$ ($a$ = $A$, $B$) in the $xy$-plane pointing from a nontopological region to a topological region, e.g., $\phi^A = 11 \pi / 6$ and $\phi^B = 5 \pi / 6$ in Fig.~\ref{fig:exchange}(a). The other parameter is the exchange-field direction $\hat{\mathbf{b}}^a = \cos \eta^a \, \hat{\mathbf{z}} + \sin \eta^a \, \hat{\mathbf{w}}^a$ acting on a MF, e.g., $\eta^A = - \pi / 2$ and $\eta^B = \pi / 2$ in Fig.~\ref{fig:exchange}(a). The state $(\phi, \eta)$ of each MF evolves under the exchange process while making a trajectory on the parameter space which is topologically a torus. When the Hamiltonian goes back to the original form after the exchange, the combined path of trajectories of two MFs forms a closed curve on the torus. Figure~\ref{fig:parameter}(a) and (b) depict the combined curves of the parameters of our process and \textcite{AliceaNP2011}'s process, respectively. Both curves wind the torus once along the angle $\phi$: they are topologically equivalent. Thus the two processes should exhibit the same exchange statistics \cite{HalperinPRB2012}.

\emph{Exact solutions for MFs.}|In the strong SOC regime, $E_\text{so} = m \alpha^2 / \hbar^2 \gg \max ( |\mathbf{M}|, \Delta)$, we can use the BdG Hamiltonian density linearized at zero momentum:
\begin{equation}
\mathcal{H}^\text{lin} = - i \alpha \partial_x \sigma_2 \tau_3 + M_x (x) \sigma_1 + M_z (x) \sigma_3 + \Delta \tau_1,
\label{eq:H-i}
\end{equation}
where the chemical potential is set to zero \cite{OregPRL2010, KlinovajaPRB2012}. In the following discussions, we set $\alpha = 1$.

To obtain an intuitive idea of the existence of MFs bound to a DW, let us neglect the exchange field $M_z$ for the moment. The Hamiltonian density $\mathcal{H}^\text{lin}$, then, can be classified as the BDI class with the particle-hole symmetry operator $\sigma_2 \tau_2 K$, the BDI time reversal symmetry operator $\sigma_1 \tau_1 K$ (different from the conventional one), and the BDI chiral symmetry operator $\sigma_3 \tau_3$ \cite{SchnyderPRB2008, *TewariPRL2012}. The Hamiltonian density $\mathcal{H}^\text{lin}$ and the BDI time reversal symmetry operator can be block-diagonalized simultaneously by employing Majorana operators instead of fermion operators. Define four Hermitian Majorana operators $\hat{\gamma}^{A,B}_{\uparrow, \downarrow}$ in such a way that
\begin{equation}
\hat{\psi}_\downarrow = (\hat{\gamma}^A_\downarrow + i \hat{\gamma}^B_\downarrow) / \sqrt{2}, \quad \hat{\psi}_\uparrow = (\hat{\gamma}^B_\uparrow + i \hat{\gamma}^A_\uparrow) / \sqrt{2}.
\end{equation}
Majorana operators are labeled by $A$ and $B$ for the convenience of the following discussion. The quasiparticle creation operator is
$
\hat{\gamma}^+ = \int dx \sum_{\substack{\alpha = \uparrow, \downarrow \\ \beta = A, B}} \left[ u^\beta_\alpha (x) \hat{\gamma}^\beta_\alpha (x) \right],
$
where the new 4-component spinor $\tilde{\Psi} = (u^A, u^B)^{\tr} \equiv (u^A_\uparrow, u^A_\downarrow, u^B_\uparrow, u^B_\downarrow)^{\tr}$ is related to the original spinor $\Psi$ by the unitary transformation $\tilde{\Psi} = U \Psi$. In this new basis, the 4-component Schr\"odinger equation $i \partial_t \tilde{\Psi} = (U \mathcal{H}^\text{lin} U^\dagger) \tilde{\Psi}$ is cast as two Dirac equations
\begin{subequations}
\label{eq:Dirac}
\begin{align}
i \gamma^\mu \partial_\mu u^A + [ - M_x (x) - \Delta ] u^A &= 0, \\
i \gamma^\mu \partial_\mu u^B + [ M_x (x) - \Delta ] u^B &= 0,
\end{align}
\end{subequations}
where we have taken the representation of Dirac matrices to be $\gamma^0 = - \sigma_2$ and $\gamma^1 = \pm i \sigma_3$ ($-$ for $u^A$). The masses of A and B, $m^A (x) \equiv - M_x (x) - \Delta$ and $m^B (x) \equiv M_x (x) - \Delta$, are essentially the topological gap: $E_g(x) = m^A (x)$ when $M_x (x) > 0$ and $E_g (x) = m^B (x)$ when $M_x (x) < 0$. These Dirac equations coincide with that of solitons with fermion number $1/2$ that \textcite{JackiwPRD1976} studied. Each Dirac equation always has a unique static normalizable solution when its mass changes sign between two ends $x \rightarrow \pm \infty$. The exchange field $\mathbf{M} = [ M_1 \tanh(x / \lambda) + M_2 ] \hat{\mathbf{x}}$ yields two zero-energy solutions \footnote{The Dirac equations~(\ref{eq:Dirac}) with the exchange field $\mathbf{M} =[M_1 \tanh(x / \lambda) + M_2]\hat{\mathbf{x}}$ are exactly solvable for an arbitrary energy with the aid of supersymmetric quantum mechanics \cite{CooperPR1995}, e.g., all the bound and continuum states have been worked out for the case of $M_2 = 0$ in Ref.~\cite{CharmchiPRD2014}.}
\begin{subequations}
\label{eq:u-soln}
\begin{align}
\hat{\gamma}^A &= \int dx \,e^{- (M_2+\Delta) x} \sech^{M_1 \lambda} (x/\lambda) \hat{\gamma}^A_\uparrow (x), \label{eq:u-soln-a} \\
\hat{\gamma}^B &= \int dx \, e^{-(M_2-\Delta) x} \sech^{M_1 \lambda} (x/\lambda) \hat{\gamma}^B_\uparrow (x)
\end{align}
\end{subequations}
up to the normalization factor. $\hat{\gamma}^A$ and $\hat{\gamma}^B$ are normalizable MF solutions when $M_2 < M_1 - \Delta$ and $M_2 < M_1 + \Delta$, respectively.

When the uniform field $M_2$ is smaller than $M_1 - \Delta$, both masses $m^{A}(x)$ and $m^B(x)$ cross zero at $\lambda \tanh^{-1}[ -(M_2 \pm \Delta) / M_1 ]$ ($+$ for A) and thus two MFs exist \cite{FuPRL2009}. Perturbations breaking the BDI time reversal symmetry split the degeneracy. First, the kinetic energy $\left( \hbar^2 \partial_x^2 / 2 m \right) \tau_3$ omitted from the original BdG Hamiltonian density~(\ref{eq:H-BdG}) hybridizes MFs with the energy $\lesssim \hbar^2 \max(M_1^2, \Delta^2) / 2 m \alpha^2$, which is finite, yet still much smaller than the bulk gap $\sim M_1, \Delta$. Second, note that both MF operators are made of $\uparrow$-spin fermion operators \cite{SticletPRL2012}. The physical fermion $\hat{f} \equiv (\hat{\gamma}^A + i \hat{\gamma}^B) / 2$ constructed from them, thus, couples to the exchange field in the $z$ direction. For $M_z (x) = M_z \sech (x / \lambda)$, the coupling energy of the fermion $\hat{f}$ to the exchange field, i.e., the hybridization energy of MFs, is proportional to $M_z \lambda$. An infinitely abrupt wall $\lambda \rightarrow 0$ reproduces the case of the unidirectional exchange field.

\begin{figure}
\includegraphics[width=0.49\columnwidth]{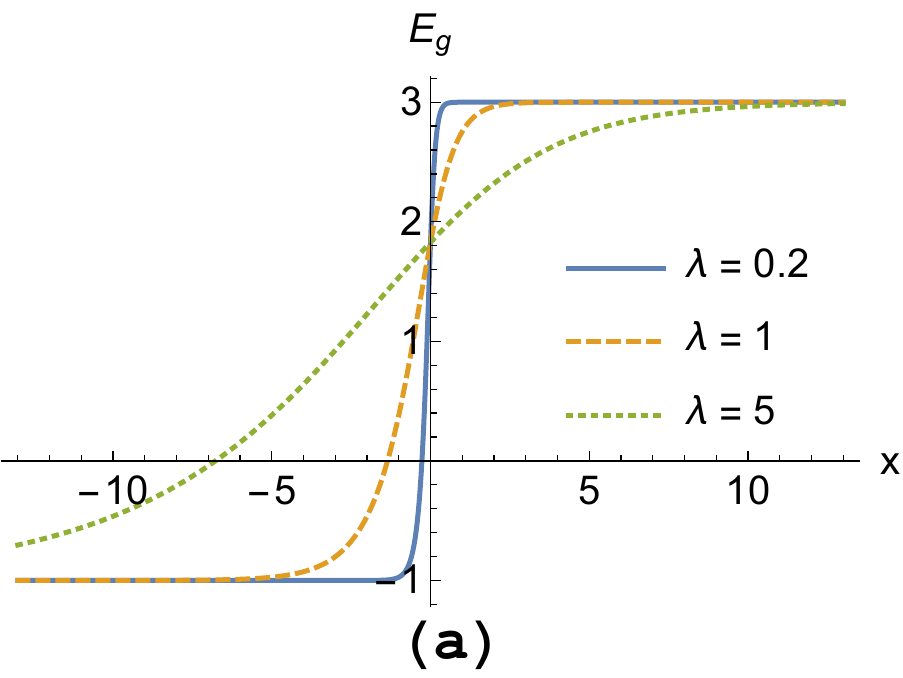}
\includegraphics[width=0.49\columnwidth]{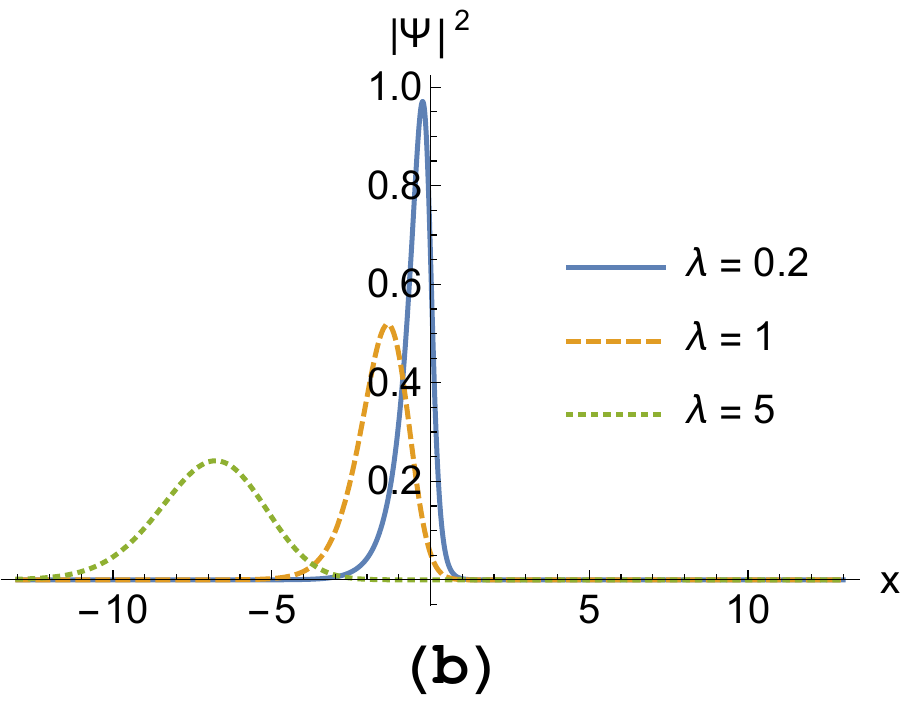}
\caption{(Color online) (a) The topological gap $E_g(x)$~(\ref{eq:E-g}) for the exchange field $\mathbf{M} = [ 2 + 2 \tanh(x / \lambda) ] \hat{\mathbf{x}} + \sech(x / \lambda) \hat{\mathbf{z}}$ with domain wall widths $\lambda = $ 0.2, 1, and 5. (b) The amplitude squared $|\Psi (x)|^2 = \left| u^A_\downarrow (x) \right|^2 + \left| u^B_\uparrow (x) \right|^2$ of MFs in Eq.~(\ref{eq:u-soln-2}), which are located most likely where the gap vanishes.}
\label{fig:soln}
\end{figure}

When the uniform field exceeds $M_1 - \Delta$, while still smaller than $M_1 + \Delta$, the left part of the wire $x \ll - \lambda$ is driven into the nontopological phase, destroying MF A. The surviving MF B is protected from perturbations, e.g., finite $M_z$, as long as the bulk gap remains finite. In particular, for the exchange field $\mathbf{M}(x)$~(\ref{eq:M}) with $M_1 = M_2 > \Delta / 2$, one MF is located at $x_0 = \lambda \tanh^{-1} (\Delta^2 / 2 M_1^2 - 1)$. The anticommutativity of the particle-hole transformation operator $\sigma_2 \tau_2 K$ and the Hamiltonian density $\mathcal{H}^\text{lin}$~(\ref{eq:H-i}) allows us to obtain a nondegenerate MF solution \cite{SauPRL2010}, given by
\begin{subequations}
\label{eq:u-soln-2}
\begin{align}
u^B_\uparrow (x) &= \frac{e^{\Delta x}}{ \sqrt[4]{1 + e^{2 x / \lambda}}} P^{-\nu}_{-1/2} \left( \frac{1}{\sqrt{1 + e^{- 2 x / \lambda}}} \right), \\
u^A_\downarrow (x) &= - (\nu + 1/2) \frac{e^{\Delta x}}{ \sqrt[4]{1 + e^{2 x / \lambda}}} P^{-\nu-1}_{-1/2} \left( \frac{1}{\sqrt{1 + e^{- 2 x / \lambda}}} \right)
\end{align}
\end{subequations}
up to the common normalization factor, where $P_\alpha^\beta (x)$ is the associated Legendre function of the first kind with degree $\alpha$ and order $\beta$ and $\nu = 2 M_1 \lambda - 1/2$ is the characteristic degree of our problem. Figure~\ref{fig:soln} shows the profile of the topological gap and the amplitude squared of the MF solution. The relative magnitude of the two components can be obtained in two limiting cases: $|u^A_\downarrow(x)| / |u^B_\uparrow(x)| \simeq 2 M_1 \lambda$ when $M_1 \lambda \ll 1$, and $\simeq 1$ when $M_1 \lambda \gg 1$. For an abrupt wall $\lambda \rightarrow 0$, the solution (\ref{eq:u-soln-2}) converges to the solution (\ref{eq:u-soln-a}) which was obtained in the absence of $M_z$.

\emph{Discussion.}|We have proposed to bind MFs to magnetic DWs in the heterostructure of a SOC nanowire, an $s$-wave superconductor, and ferromagnet wires. We have also studied MFs analytically in the strong spin-orbit regime. A topologically stable DW provides a robust matrix for strong MFs. Typical strong SOC semiconducting nanowires (e.g., InAs), magnetic insulators (e.g., EuO), and $s$-wave superconductors (e.g., Nb) provide the parameters $\alpha \sim 5 \, \text{meV nm}, \, |\mathbf{M}| \sim 1 \, \text{meV}, \, \Delta \sim 0.5 \, \text{meV}$ \cite{SauPRL2010}, which must be sufficient to bind MFs to DWs. Static MFs at DWs can be experimentally observed by measuring the zero-bias peak \cite{MourikScience2012, *DasNP2012, *FinckPRL2013}, free from spurious ``end'' effects such as localized states near the wire's end. 

Diverse means to control DW motion allow us to manipulate MFs. In particular, we have proposed a process braiding two MFs in the $Y$ junction performed by thermally-driven DW motion, which exhibits non-Abelian exchange statistics. Thermally-driven motion of a DW has been observed in yttrium iron garnet films \cite{JiangPRL2013}: the DW moves at the velocity $v \sim 100 \, \mu \text{m/s}$ for a temperature gradient $\nabla T \sim 2 \, \mu \text{eV} / \mu \text{m}$. The resultant temperature drop over the DW width $\lambda \sim 60 \, \text{nm}$ is much smaller than the induced topological gap $\sim 200 \, \mu \text{eV}$ \cite{MourikScience2012}. 

A domain wall in an easy-axis magnetic wire is a one-dimensional topological soliton. It would be worth pursuing higher dimensional generalizations of our proposal. For example, a thin film of Fe$_{0.5}$Co$_{0.5}$Si subject to an external field supports a lattice of skyrmions which are two-dimensional topological solitons \cite{YuNature2010}. For a single skyrmion, the local direction of the magnetization changes smoothly from $+ \hat{\mathbf{z}}$ at its center to $- \hat{\mathbf{z}}$ at its outside (or vice versa). The magnetization on its radial line defines a domain wall located at its edge where the magnetization is in-plane. Skyrmions, therefore, may host MFs on their edges under suitable conditions, which can be braided by thermally-induced motions of skyrmions \cite{KongPRL2013}.

\begin{acknowledgments}
We thank Jelena Klinovaja, Rahul Roy, Scott Bender, So Takei, and Thomas Jackson for useful discussions. This work was supported by the US DOE-BES under Award No. $\text{DE-SC0012190}$ and in part by FAME (an SRC STARnet center sponsored by MARCO and DARPA), the ARO under Contract No. 911NF-14-1-0016 (S.K.K. and Y.T.), and AFOSR FA9550-13-1-0045 (S.T.).
\end{acknowledgments}

\bibliographystyle{apsrev4-1-nourl}
\bibliography{MF}

\end{document}